\documentclass{LMCS}

\usepackage{amssymb}
\usepackage[latin1]{inputenc}
\usepackage{mathpartir}
\usepackage{amsmath}
\usepackage{enumerate,hyperref}

\newcommand{\NN}{\mathbb{N}}

\newcommand{\UM}{A}
\newcommand{\EM}{E}
\newcommand{\mneg}{\dot\neg}
\newcommand{\mand}{\mathop{\dot\land}}
\newcommand{\mor}{\mathop{\dot\lor}}

\newcommand{\nm}[1]{\dot{#1}}

\newcommand{\Dis}{D}

\newcommand{\Rand}{\mathcal{R}_{\mand}}
\newcommand{\Ror}{\mathcal{R}_{\mor}}
\newcommand{\Rbox}{\mathcal{R}_{\Box}}
\newcommand{\Rdmnd}{\mathcal{R}_{\Diamond}}
\newcommand{\Rum}{\mathcal{R}_{\UM}}
\newcommand{\Rem}{\mathcal{R}_{\EM}}
\newcommand{\RN}{\mathcal{R}_{N}}
\newcommand{\RNn}{\mathcal{R}_{\bar N}}
\newcommand{\RNnI}{\mathcal{R}_{\dneq}^\bot}
\newcommand{\Rneg}{\mathcal{R}_{\mneg}^\bot}

\newcommand{\Nam}{\mathcal{N}}
\newcommand{\Ver}{\mathrm{S}}
\newcommand{\Bool}{\mathrm{B}}
\newcommand{\deq}{{\doteq}}
\newcommand{\dneq}{{\not\doteq}}
\newcommand{\Dmnd}[2]{\langle #1\rangle_{#2}}
\newcommand{\Bx}[2]{[#1]_{#2}}

\newcommand{\Cl}{\Gamma}
\newcommand{\Dl}{\Delta}
\newcommand{\Clx}{\tilde\Gamma}
\newcommand{\Dlx}{\tilde\Delta}
\newcommand{\Acc}[3]{#2\vartriangleright_\Cl^{#1}#3}
\newcommand{\AccRef}[3]{#2\trianglerighteq_\Cl^{#1}#3}
\newcommand{\AccRefCli}[4]{#3\trianglerighteq_{#1}^{#2}#4}
\newcommand{\AccDl}[3]{#2\vartriangleright_\Dl^{#1}#3}
\newcommand{\AccRefDl}[3]{#2\trianglerighteq_\Dl^{#1}#3}
\newcommand{\Ref}{R}
\newcommand{\Trans}{T}
\newcommand{\Pat}{\mathrm{Pat}\,}
\newcommand{\PatCl}[2]{P_\Cl^{#1}#2}
\newcommand{\PatDl}[2]{P_\Dl^{#1}#2}
\newcommand{\UPatCl}[1]{P_\Cl #1}

\newcommand{\PatCli}[3]{P_{#1}^{#2}#3}

\newcommand{\Rt}{\mathcal{R}_{\Trans}}

\newcommand{\mci}{\mathcal{I}}
\newcommand{\Sta}{\mathrm{S}}
\newcommand{\GMod}{\mci}
\newcommand{\ModCl}{\mci}
\newcommand{\STer}{\mathcal{S}}
\newcommand{\STerNew}{\STer'}
\newcommand{\Rel}{\mathrm{Rel}\,}
\newcommand{\ext}[1]{\stackrel{#1}{\to}}
\newcommand{\fE}{\psi_\EM}
\newcommand{\fD}{\psi_\Diamond}
\newcommand{\subsetax}{\sqsubseteq}
\newcommand{\evClo}[1]{\hat #1}

\newcommand{\Calc}{\mathcal{T}}
\newcommand{\CalcX}{\mathcal{T}_\subsetax}

\def\doi{7 (1:5) 2011}
\lmcsheading%
{\doi}
{1--21}
{}
{}
{May\phantom.~14, 2010}
{Mar.~21, 2011}
{}

\begin{document}

\title[Terminating Tableaux for Graded Hybrid Logic]
{Terminating Tableaux for Graded Hybrid Logic with
  Global Modalities and Role Hierarchies}
\author[M.~Kaminski]{Mark Kaminski}
\address{Saarland University, Saarbrücken, Germany}
\email{\{kaminski,smolka\}@ps.uni-saarland.de}
\author[S. Schneider]{Sigurd Schneider}
\address{\vskip-6 pt}
\author[G. Smolka]{Gert Smolka}
\address{\vskip-6 pt}
\titlecomment{A preliminary version of this work appeared in~\cite{KaminskiEtAlTableaux09}.}
\begin{abstract}
  We present a terminating tableau calculus for graded hybrid logic with global
  modalities, reflexivity, transitivity and role hierarchies.
  Termination of the system is achieved through pattern-based blocking. Previous
  approaches to related logics all rely on chain-based blocking.
  Besides being conceptually simple and suitable for efficient implementation,
  the pattern-based approach gives us a \textsc{NExpTime} complexity
  bound for the decision procedure.
\end{abstract}
\maketitle
\section{Introduction}
Graded modal logic~\cite{Fine72} is a powerful generalization of basic modal
logic.
Most prominently, graded modalities are used in
description logics, rich modal languages tailored for knowledge representation
that have a wide range of practical
applications~\cite{HandbookDL}.
Graded modal logic
allows to constrain the number
of accessible states satisfying a certain property. So, the modal formula
$\Diamond_n p$ is true in a state $x$ if $x$ has at least $n+1$ successors
satisfying $p$. Analogously to ordinary modal logic, graded modal logic can be
extended by nominals~\cite{ArecesTenCateHandbook07}.
The resulting language, graded hybrid logic, can be extended further by adding
global modalities~\cite{GorankoPassy92},
which allow to specify properties that are to
hold in all states.

Role hierarchies were first studied by Horrocks~\cite{HorrocksPHD} in the
context of description logics. Using inclusion assertions of the form
$r\subsetax r'$, one can specify that the role (relation) $r$ is contained in
the role $r'$. Role hierarchies are of particular interest when considered
together with
transitivity assertions for roles~\cite{Sattler96,BaaderLutzHandbook07}.
The description logic $\mathcal{SHOQ}$~\cite{HorrocksSattler01} combines
the expressive means provided by nominals, graded modalities, role hierarchies
and transitive roles.

We present a terminating tableau calculus for graded multimodal logic extended
by nominals, global modalities, reflexive and transitive roles, and
role hierarchies.
The modal language under consideration in the present work is equivalent to
$\mathcal{SHOQ}$
extended by reflexive roles and a universal role, both extensions also
being known from $\mathcal{SROIQ}$~\cite{HorrocksEtAl06}.

The most important difference of our approach to existing calculi for
$\mathcal{SHOQ}$ and stronger
logics~\cite{HorrocksSattler01,HorrocksSattler07,HorrocksEtAl06} is
the technique used to achieve termination of the tableau construction. The
established tableau algorithms all rely on modifications of Kripke's
chain-based blocking technique~\cite{Kripke1963}. Chain-based blocking assumes
a precedence order
on the nominals (also known as 
nodes or prefixes)
of a tableau branch, and prevents processing of nominals that are subsumed by
preceding nominals.
In the simplest case, the precedence order is chosen
to be the ancestor relation among nominals (ancestor blocking).
In general, however, it may be any order that contains the ancestor relation
(anywhere
blocking~\cite{BaaderEtAl96,MotikEtAl07}).
Ancestor blocking gives an exponential bound on the length of ancestor chains,
resulting in a double exponential bound on the size of tableau branches.
Depending on the choice of the precedence order,
anywhere blocking can lower this bound to
a single exponential. However, the size bound on tableau branches does not seem
to translate easily to a complexity bound for the decision procedures
in~\cite{HorrocksSattler01,HorrocksSattler07,HorrocksEtAl06}
(\cite{HorrocksSattler01,HorrocksSattler07} show a \textsc{2-NExpTime} bound,
while \cite{HorrocksEtAl06} leaves complexity open). We feel that
the main difficulty in
obtaining better complexity bounds is the algorithms being non-cumulative.

A tableau system is called cumulative if its rules never update or delete
formulas. In contrast to most systems in the literature, calculi devised
for description logics are often not cumulative.
By giving up cumulativity, it is possible to
obtain a more direct
correspondence between tableau branches and the candidate models they represent.
So, for instance, a non-cumulative calculus may merge several nominals into one
if the nominals are found to be semantically equivalent.
In this way,
one can achieve that every state of a candidate model is represented by
exactly one nominal.
This close correspondence
is intuitive and may simplify model existence arguments.
At the same time,
non-cumulative rules are typically more complex than their cumulative
counterparts, which may complicate the presentation of a calculus. More
importantly, cumulative systems are usually more amenable to termination
and complexity analysis.
The problem with non-cumulative systems is that rules that can update or delete
formulas may potentially undo earlier changes made to a tableau branch.
For instance, consider two tableau branches $\Cl$
and $\Dl$, where $\Dl$ is obtained from $\Cl$ by some sequence of
tableau rule applications. In a non-cumulative calculus, it is conceivable
that by applying some rule to $\Dl$, we may obtain $\Cl$ again. Clearly,
such a calculus is non-terminating even if the size of tableau branches can
be bounded. Often, termination of non-cumulative calculi can only be achieved
if rule application follows some fixed
strategy~\cite{BaaderSattler01,HorrocksSattler07,HorrocksEtAl06}.
And even then, size bounds on tableau branches do not immediately yield
time complexity bounds. To construct a branch of size $n$,
a non-cumulative system may need significantly more than $n$ rule
applications.
Cumulative calculi, on the other hand, are guaranteed to enlarge the branch by
at least one formula in every step. Therefore, a size bound on tableau branches
can immediately be interpreted as an upper bound on the non-deterministic time
complexity of the decision procedure.

Unlike~\cite{HorrocksSattler01,HorrocksSattler07,HorrocksEtAl06},
our calculus is cumulative. Cumulativity of the calculus in the presence of
nominals is achieved following~\cite{KaminskiSmolkaJoLLI} by representing
equality constraints via an equivalence relation on nominals.
Termination of our system is achieved through pattern-based
blocking~\cite{KaminskiSmolkaM4M07,KaminskiSmolkaJoLLI}. Pattern-based blocking
is conceptually simpler
than chain-based techniques in that it does not need an order on the nominals,
and seems promising as it comes to
efficient implementation~\cite{GoetzmannEtAl:2009:Spartacus}.
Pattern-based blocking provides an exponential bound on the size of tableau
branches and on the number of tableau rule applications for a single branch.
Thus it limits the complexity of the associated decision procedure to
\textsc{NExpTime}.
To deal with graded modalities, we extend the blocking conditions
in~\cite{KaminskiSmolkaM4M07,KaminskiSmolkaJoLLI}, preserving the exponential
size bound on tableau branches.

It is worth noting that, despite of the close interplay between pattern-based
blocking and abstract representation of state equality in the present work,
the two techniques should be seen as independent and applicable in isolation
from each other. In fact, pattern-based blocking
was introduced in~\cite{KaminskiSmolkaM4M07} for a non-cumulative system where
equality was treated by means of a substitution operation on branches. Also, in
previous work~\cite{KaminskiSmolkaJoLLI}, we show how abstract treatment of
equality can be combined with chain-based blocking to obtain cumulative,
terminating tableau calculi for hybrid logic with converse modalities and the
difference modality.

We begin by presenting a calculus for graded hybrid logic with global
modalities. We argue that the blocking conditions used
in~\cite{KaminskiSmolkaM4M07,KaminskiSmolkaJoLLI} are insufficient in the
presence of graded modalities. We extend pattern-based blocking to account
for the increased expressive power and argue the completeness and termination
of the resulting calculus. In the second part of the paper, we extend our
calculus further by allowing reflexivity, transitivity and inclusion assertions.
It turns out that in the presence of inclusion assertions, the blocking
condition used for the basic calculus needs to be extended once again.

\section{Graded Hybrid Logic with Global Modalities and
  Role Inclusion}
Following~\cite{KaminskiSmolkaIJCAR08,KaminskiSmolkaJoLLI},
we represent modal logic in simple type theory (see,
e.g.,~\cite{Farmer08,BrownSmolka10}).
This way we can make use of a rich syntactic and semantic framework and modal
logic does not appear as an isolated formal system.
We start with two base types $\Bool$ and $\Ver$.  The
interpretation of $\Bool$ is fixed and
consists of two
truth values.  The interpretation of $\Ver$ is a
nonempty set whose elements are called \emph{worlds} or
\emph{states}.  Given two types $\sigma$ and $\tau$, the
\emph{functional type} $\sigma\tau$ is interpreted as
the set of all total functions from the interpretation
of $\sigma$ to the interpretation of $\tau$.  We write
$\sigma_1\sigma_2\sigma_3$ for
$\sigma_1(\sigma_2\sigma_3)$.

We assume a countable set of \emph{names}, which we partition into a
countable set of \emph{variables} and a set of \emph{constants}.
We employ three kinds of variables: \emph{Nominal
  variables\/} $x$, $y$, $z$ of type $\Ver$,
\emph{propositional variables} $p$, $q$ of type
$\Ver\Bool$, and \emph{role variables} $r$ of
type $\Ver\Ver\Bool$.  Nominal variables are called
\emph{nominals} for short, and role variables are called \emph{roles}.
We assume there are infinitely many nominals.
We use the logical constants
\begin{align*}
  \bot,\top & : \Bool &
  \neg & : \Bool\Bool &
  \lor,\land,\to & :  \Bool\Bool\Bool &
  \deq & : \Ver\Ver\Bool &
  \exists,\forall & : (\Ver\Bool)\Bool
\end{align*}
Terms are defined as usual.  We write $s t$ for
applications, $\lambda{x}.s$ for abstractions, and
$s_1 s_2 s_3$ for $(s_1 s_2)s_3$.  We also use infix
notation, e.g., $s\land{t}$ for $(\land)st$.

Terms of type $\Bool$ are called \emph{formulas}.  We
employ some common notational conventions:
$\exists x.s$ for $\exists(\lambda{x}.s)$, \ $\forall x.s$ for
$\forall(\lambda{x}.s)$, and $x\dneq y$ for
$\neg(x\deq{y})$.
Given a set $X$ of nominals, we use the following abbreviation:
\begin{align*}
  \Dis X &~:=~\bigwedge_{\begin{subarray}{c}x,y\in X\\x\ne y\end{subarray}}x\dneq y
\end{align*}

We use the following constants:
\begin{align*}
  \subsetax&:(\Ver\Ver\Bool)(\Ver\Ver\Bool)\Bool
  & r_1\subsetax r_2 &~=~\forall x y.r_1 x y\to r_2 x y\\
  \Ref&:(\Ver\Ver\Bool)\Bool
  & \Ref r &~=~\forall x.r x x\\
  \Trans&:(\Ver\Ver\Bool)\Bool
  & \Trans r &~=~\forall x y z.r x y\land r y z\to r x z
\end{align*}
To the right of each constant is an equation defining its semantics.
We call formulas of the form $r\subsetax r'$ \emph{(role) inclusion assertions}.
Formulas $\Ref r$ and $\Trans r$ are called \emph{reflexivity} and
\emph{transitivity assertions}, respectively.

We write $\exists^n X.s$ for
$\exists x_1\dots x_n.s$ if $|X|=n$ and $X=\{x_1,\ldots,x_n\}$.
The \emph{modal constants} are then defined as follows:
\begin{align*}
  \mneg&:(\Ver\Bool)\Ver\Bool
  &\mneg p x &~=~\neg(p x) \\
  \mand&:(\Ver\Bool)(\Ver\Bool)\Ver\Bool
  &(p\mand q)x &~=~p x\land{q x} \\
  \mor&:(\Ver\Bool)(\Ver\Bool)\Ver\Bool
  &(p\mor q)x &~=~p x\lor{q x} \\
  \Dmnd{\_}{n}&:(\Ver\Ver\Bool)(\Ver\Bool)\Ver\Bool
  &\Dmnd{r}{n}p x &~=~\exists^{n+1} Y.\,
  \Dis Y\land(\textstyle{\bigwedge_{y\in Y}r x y\land p y})\\
  \Bx{\_}{n}&:(\Ver\Ver\Bool)(\Ver\Bool)\Ver\Bool
  &\Bx{r}{n}p x &~=~\forall^{n+1} Y.\,
  (\textstyle{\bigwedge_{y\in Y}r x y})\land\Dis Y\to\textstyle{\bigvee_{y\in Y}p y}\\
  \EM_n&:(\Ver\Bool)\Ver\Bool
  & \EM_n p x &~=~\exists^{n+1} Y.\,
  \Dis Y\land\textstyle{\bigwedge_{y\in Y}p y}\\
  \UM_n&:(\Ver\Bool)\Ver\Bool
  & \UM_n p x &~=~\forall^{n+1} Y.\,
  \Dis Y\to\textstyle{\bigvee_{y\in Y}p y} \\
  \nm{\_}&:\Ver\Ver\Bool
  & \nm{x}y &~=~x\deq y\\[1mm]
&\!\!\!\!\!\!\!\!\!\!{~}\makebox[2mm][l]{\footnotesize{\mbox{\,where $n\ge 0$ in all equations}}}
\end{align*}
The semantics of boxes and diamonds is defined following
\cite{FattorosiBarnabaDeCaro85,vanDerHoekDeRijke95,OhlbachEtAl96}.
Intuitively, it can be described as follows:
\begin{description}
\item[$\EM_n p$] There are at least $n+1$ states satisfying $p$.
\item[$\UM_n p$] All states but possibly $n$ exceptions satisfy $p$.
\item[$\Dmnd{r}{n} p$] There are at least $n+1$ $r$-successors satisfying $p$.
\item[$\Bx{r}{n} p$] All $r$-successors but possibly $n$ exceptions satisfy
  $p$.
\end{description}
In accordance with the usual modal intuition, ``formulas'' of modal logic
are seen as predicates of type $\Ver\Bool$ denoting sets of states.
They can be represented as \emph{modal expressions}
according to the following grammar:
\begin{align*}
  t &~::=~ p\;|\;\nm x\;|\;\mneg t\;|\;t\mand t\;|\;t\mor t\;|\;\Dmnd{r}{n}t\;
  |\;\Bx{r}{n}t\;|\;\EM_n t\;|\;\UM_n t
\end{align*}

As with the propositional connectives, we use infix notation for $\mand$ and
$\mor$. Unlike with the propositional connectives, we assume the
application of modal operators to have a higher precedence than regular
functional application.
So, for instance, we write $\mneg\Dmnd{r}{2}\nm y\mor p~x$ for
$((\mneg(\Dmnd{r}{2}(\nm y)))\mor p)x$.

An \emph{interpretation} is a function $\mci$ mapping $\Bool$ to the set
$\{0,1\}$, $\Ver$ to a non-empty set, a functional type $\sigma\tau$ to the
set of all total functions from $\mci\sigma$ to $\mci\tau$, and
every name $x:\sigma$ to an element of
$\mci\sigma$ (i.e., $\mci x\in\mci\sigma$) such that the logical constants get
their usual meaning:
\begin{align*}
  \GMod\bot=0&\,\,\,\,\,\textup{and}\,\,\,\,\,\GMod\top=1 &(\GMod\neg)a=1&\,\iff\,a=0\\
  (\GMod\land)a b=1&\,\iff\,a=1\textup{ and }b=1
  &(\GMod\lor)a b=1&\,\iff\,a=1\textup{ or }b=1\\
  (\GMod\to)a b=1&\,\iff\,a=0\textup{ or }b=1
  &(\GMod\deq)a b=1&\,\iff\, a=b\\
  (\GMod\exists)f=1&\,\iff\,\textup{$f a=1$ for some $a\in\mci\Ver$}~~
  &(\GMod\forall)f=1&\,\iff\,\textup{$f a=1$ for all $a\in\mci\Ver$}
\end{align*}
If $\mci$ is an interpretation, $x:\sigma$ is a variable, and $a\in\mci\sigma$,
then $\mci^x_a$ denotes the interpretation
that agrees everywhere with $\mci$ but possibly on $x$ where it yields $a$.
Every interpretation $\mci$ can be extended to a function $\hat\mci$
that maps every term $s:\sigma$ to an element of $\mci\sigma$
such that:
\begin{align*}
  \hat\mci x&~=~\mci x\\
  \hat\mci(s t)&~=~(\hat\mci s)(\hat\mci t)\\
  \hat\mci(\lambda x.s)&~=~
  \{(a,\,\widehat{\mci^x_a}s)\,|\,a\in\mci\sigma\}
  \qquad\qquad~ 
  \textup{if $x:\sigma$}
\end{align*}
Since $\hat\mci$ is uniquely determined by $\mci$, in the following we write
$\mci s$ for $\hat\mci s$ for convenience.
A \emph{modal interpretation} is an
interpretation that, in addition, satisfies the above equations defining the
constants $\subsetax$, $\Ref$, $\Trans$,
$\mneg$, $\mand$, $\mor$, $\Dmnd{\_}{n}$, $\Bx{\_}{n}$, $\EM$, $\UM$, 
$\nm{\_}\,$.
If \mbox{$\GMod s=1$}, we say that $\GMod$
\emph{satisfies}\/ $s$, or that $\GMod$ is a \emph{model} of $s$.
A modal interpretation $\GMod$ satisfies a set $\Cl$ of formulas
($\GMod$ is a \emph{model} of $\Cl$) if $\GMod$ satisfies every
formula in $\Cl$.
A formula (a set of formulas) is called \emph{satisfiable} if it
has a model.

\section{Graded Hybrid Logic with Global Modalities}
We begin with a tableau calculus for the restricted language without
inclusion, reflexivity or transitivity assertions.

\subsection{Branches}
For the sake of simplicity, we define our tableau calculus on negation normal
expressions, i.e., terms of the form:
\begin{align*}
  t &~::=~ p\;|\;\mneg p\;|\;\nm x\;|\;\mneg\nm x\;|\;t\mand t\;|\;t\mor t\;
  |\;\Dmnd{r}{n}t\;|\;\Bx{r}{n}t\;|\;\EM_n t\;|\;\UM_n t
\end{align*}

A \emph{branch} $\Cl$ is a finite set of formulas $s$ of the form
\begin{align*}
  s &~::=~ t x\;|\;r x y\;|\;x\deq y\;|\;x\dneq y\;|\;\bot
\end{align*}
where $t$ is a negation-normal modal expression of the above form.
Formulas of the form $r x y$ are called
\emph{accessibility formulas} or \emph{edges}.
We use the formula $\bot$ to explicitly mark unsatisfiable branches.
We call a branch $\Cl$ \emph{closed} if $\bot\in\Cl$. Otherwise, $\Cl$ is
called \emph{open}.
The branch consisting of the initial formula (or formulas) to be tested for
satisfiability is called the \emph{initial branch}.

Let $\Cl$ be a branch.
With $\sim_\Cl$ we denote the least equivalence relation $\sim$ on nominals such
that $x\sim y$ for every equation $x\deq y\in\Cl$.
Let $R(x,y)$ denote a term of the form $x\deq y$, $x\dneq y$, or $r x y$.
We define the \emph{equational closure $\Clx$} of a branch $\Cl$ as
\begin{align*}
\Clx&~:=~\Cl\cup\{t x\,|\,t\textup{ modal expression}~\land~\exists x':~x'\sim_\Cl x~\land~t x'\in\Cl\}\\
&\phantom{~:=~\Cl\;}\cup\{R(x,y)\,|\,\exists x',\,y':~x'\sim_\Cl x~\land~
y'\sim_\Cl y~\land~R(x',y')\in\Cl\}
\end{align*}
Note that for all nominals $x$ and $y$, $x\sim_\Cl y$ holds if and only if
$x\deq y\in\Clx$.
Since $\Clx$ only contains nominals, modal expressions and roles that already
occur on $\Cl$, $\Clx$ clearly is finite if $\Cl$ is finite.
Reasoning with respect to $\Clx$ can be implemented efficiently using
disjoint-set forests, as demonstrated
in~\cite{Goetzmann09,GoetzmannEtAl:2009:Spartacus}.

\subsection{Evidence}
The proof of model existence for our calculus proceeds in two stages.
Applied to a satisfiable initial branch, the rules of the calculus
(defined in Sect.~\ref{sec:basic-tableau-rules}) construct a
\emph{quasi-evident} branch (defined in Sect.~\ref{sec:control}). We
show that every quasi-evident branch can be extended to an \emph{evident}
branch. For evident branches, we show model existence.
Intuitively, we call a branch evident if it contains a complete
syntactic description of a model of all of its formulas.

We write $D_\Cl X$ as an abbreviation for
$\forall x,y\in X\colon\, x\ne y\,\Longrightarrow\, x\dneq y\in\Clx\,\lor\,
y\dneq x\in\Clx$.
A branch $\Cl$ is called \emph{evident} if it satisfies all of the following
\emph{evidence conditions}:
\begin{align*}
  (t_1\mand t_2)x\in\Cl &~\Rightarrow~ t_1 x\in\Clx\,\land\,t_2 x\in\Clx\\
  (t_1\mor t_2)x\in\Cl &~\Rightarrow~ t_1 x\in\Clx\,\lor\,t_2 x\in\Clx\\
  \Dmnd{r}{n}t x\in\Cl
  &~\Rightarrow~
  \exists^{n+1} Y\!:~D_\Cl Y\,\land\,
  \{r x y,t y\,|\,y\in Y\}\subseteq\Clx\\
  \Bx{r}{n}t x\in\Cl
  &~\Rightarrow~ |\{y\,|\,r x y\in\Clx,~t y\notin\Clx\}/_{\sim_\Cl}|\le n\\
  \EM_n t x\in\Cl &~\Rightarrow~
  \exists^{n+1} Y\!:~
  D_\Cl Y\,\land\,
  \{t y\,|\,y\in Y\}\subseteq\Clx\\
  \UM_n t x\in\Cl &~\Rightarrow~ |\{y\,|\,t y\notin\Clx\}/_{\sim_\Cl}|\le n\\
  \nm x y\in\Cl &~\Rightarrow~ x\sim_\Cl y\\
  \mneg\nm x y\in\Cl &~\Rightarrow~ x\not\sim_\Cl y \\
  x\dneq y\in\Cl &~\Rightarrow~ x\not\sim_\Cl y\\
  \neg p x\in\Cl &~\Rightarrow~ p x\notin\Clx
\end{align*}
A formula $s$ is called \emph{evident on $\Cl$} if $\Cl$ satisfies the
right-hand side of the evidence
condition corresponding to $s$. For instance, $(t_1\mand t_2)x$
is evident on $\Cl$ if and only if $\{t_1 x,t_2 x\}\subseteq\Clx$.

Given a term $t$, we write $\Nam t$ for the set of
nominals that occur in $t$.
The notation is extended to
sets of terms in the natural way:
$\Nam\Cl:=\bigcup\{\Nam t\,|\,t\in\Cl\}$.

\begin{thm}[Model Existence] \label{thm:evident-branches-sat}
  Every evident branch has a finite model.
\end{thm}

\proof 
Let $\Cl$ be an evident branch and let $x_0\in\Nam\Cl$. Let $\rho$ be a
function from finite sets of nominals to nominals such that $\rho X\in X$
whenever $X$ is nonempty. We define the interpretation $\ModCl$
such that:
\begin{eqnarray*}
  \ModCl\Sta&:=&\{\rho\{y\,|\,y\sim_\Cl x\}\,|\,x\in\Nam\Cl\}\\
  \ModCl x&:=&\textup{if }x\in\Nam\Cl\textup{ then }
  \rho\{y\in\Nam\Cl\,|\,y\sim_\Cl x\}\textup{ else }\ModCl x_0\\
  \ModCl p&:=&\{x\in\ModCl\Ver\,|\,p x\in\Clx\}\\
  \ModCl r&:=&\{(x,y)\in(\ModCl\Ver)^2\,|\,r x y\in\Clx\}
\end{eqnarray*}
Intuitively, we construct $\ModCl$ by interpreting
$\Sta$ as the quotient of the nominals on $\Cl$
by $\sim_\Gamma$, where each equivalence class is
represented by a fixed element of the class selected by $\rho$.
Nominals on $\Cl$ are mapped to their corresponding equivalence classes.
All other nominals are mapped to some arbitrary state.
Propositional variables and roles are interpreted
as the smallest sets that are consistent with the respective assertions on
$\Cl$.
Since $\Cl$ is finite by definition, so is $\ModCl$.
Note that in the last two lines of the definition, we interpret the set notation
as a convenient description for the respective characteristic functions.

We now show that, for all $s\in\Cl$, $\ModCl$ satisfies $s$ by induction on $s$.
Let $s\in\Cl$. We proceed by case analysis.
\begin{enumerate}[$\bullet$]
\item $s=p x$. Since $\ModCl x\sim_\Cl x$, we have $p(\ModCl x)\in\Clx$.
  The claim follows.
\item $s=\mneg p x$. It suffices to show that $\ModCl(p x)=0$.
  By the evidence condition for $s$, $p x\notin\Clx$. Hence
  $p(\ModCl x)\notin\Clx$. The claim follows.
\item $s=r x y$. Then $r(\ModCl x)(\ModCl y)\in\Clx$, and hence
  $(\ModCl x,\ModCl y)\in\ModCl r$.
\item $s=x\deq y$. It suffices to show that $\ModCl x=\ModCl y$, which
  is the case as $x\sim_\Cl y$ by the definition of $\sim_\Cl$.
\item $s=x\dneq y$.
  By the evidence condition for $s$, $x\not\sim_\Cl y$. Hence
  $\ModCl x\not\sim_\Cl\ModCl y$. The claim follows.
\item $s=\Dmnd{r}{n}t x$. By the evidence condition for $s$,
  there is a set $Y$ of cardinality $n+1$ such that $D_\Cl Y$ and for
  all $y\in Y$, $\{r x y,t y\}\subseteq\Clx$. By the inductive hypothesis
  for the disequations required by $D_\Cl Y$, we have
  $|Y/_{\sim_\Cl}|=|\{\ModCl y\,|\,y\in Y\}|=n+1$. By the inductive hypothesis
  for the formulas $r x y$ and $t y$ (for all $y\in Y$), we have
  $(\ModCl x,\ModCl y)\in\ModCl r$, and $\ModCl$ satisfies $t y$.
  The claim follows.
\item $s=\Bx{r}{n}t x$. By the evidence condition for $s$,
  $|\{y\,|\,r x y\in\Clx,~t y\notin\Clx\}/_{\sim_\Cl}|\le n$. Since
  $\ModCl x\sim_\Cl x$
  whenever $x\in\Nam\Cl$, we have for all $x,y\in\Nam\Cl$:
  $(\ModCl x,\ModCl y)\in\ModCl r
  ~\Leftrightarrow~r(\ModCl x)(\ModCl y)\in\Clx~
  \Leftrightarrow~r x y\in\Clx$. Hence
  $|\{y\,|\,r x y\in\Clx,~t y\notin\Clx\}/_{\sim_\Cl}|=
  |\{\ModCl y\,|\,(\ModCl x,\ModCl y)\in\ModCl r,~t y\notin\Clx\}|\le n$.
  Moreover, by the inductive hypothesis, $\ModCl$ satisfies $t y$ whenever
  $t y\in\Clx$. The claim follows.
\end{enumerate}
The cases $s=(t_1\mor t_2)x$, $s=(t_1\mand t_2)x$ are straightforward.
The cases $s=\nm x y$ and $s=\mneg\nm x y$ proceed analogously to
$s=x\deq y$ and, respectively, $s=x\dneq y$,
and the cases
$s=\EM_n t x$ and $s=\UM_n t x$ are analogous but simpler than
$s=\Dmnd{r}{n}t x$ and, respectively, $s=\Bx{r}{n}t x$.
\qed

\subsection{Tableau Rules} \label{sec:basic-tableau-rules}
The tableau rules of our basic calculus $\Calc$ are defined in
Fig.~\ref{BasicTableauRules}.
In the rules, we write $\exists x\in X:\Gamma(x)$ for
$\Gamma(x_1)~|~\dots~|~\Gamma(x_n)$, where $X=\{x_1,\ldots,x_n\}$
and $\Gamma(x)$ is a set of formulas parameterized by $x$. In case
$X=\emptyset$, the notation translates to $\bot$.
Dually, we write
$\forall x\in X:\Gamma(x)$ for $\Gamma(x_1),\ldots,\Gamma(x_n)$
($X=\{x_1,\ldots,x_n\}$).
If $X=\emptyset$, the notation stands for the empty set of formulas.

The side condition of $\Rdmnd$ uses the notion of quasi-evidence
that we will introduce in Sect.~\ref{sec:control}. For now, we assume the rule
is formulated with the restriction ``$\Dmnd{r}{n}t x$ not evident on $\Cl$''.

Note that for
$n=0$, the rules $\Rdmnd$ and $\Rbox$ instantiate, modulo obvious
simplifications, to their respective non-graded counterparts:
\begin{mathpar}
\inferrule*[right=\mbox{$y\textup{ fresh},~\Dmnd{r}{0}t x\textup{ not quasi-evident on }\Cl$}]{\Dmnd{r}{0}t x}{r x y,~t y}\and
\inferrule*[right=\mbox{$r x y\in\Clx$}]{\Bx{r}{0}t x}{t y}
\end{mathpar}
A branch $\Dl$ is called a \emph{proper extension} of a branch $\Cl$ if
$\Dl\supseteq\Cl$ and $\Dlx\supsetneq\Clx$. Note that if $\Dl$ is a proper
extension of $\Cl$, in particular it holds $\Dl\supsetneq\Cl$.
The converse does not hold: Let $\Cl:=\{\nm x y,\,x\deq z,\,z\deq y\}$
and $\Dl:=\Cl\cup\{x\deq y\}$. Then $\Dl\supsetneq\Cl$ but $\Dl$ is not
a proper extension of $\Cl$.
We implicitly restrict the applicability of the tableau rules so that
a rule $\mathcal{R}$ is only applicable to a formula $s\in\Cl$ if all of
the alternative branches $\Dl_1,\ldots,\Dl_n$ resulting from this application
are proper extensions of $\Cl$.
Moreover, we require that for every $i,j$ with $1\le i<j\le n$,
$\Dlx_i\ne\Dlx_j$. Whenever a rule produces several alternative branches whose
equational closure is equal, by the following proposition it suffices to consider
only one of them to preserve soundness.

\begin{prop}
  Let\/ $\GMod$ be a modal interpretation and\/ $\Cl$, $\Dl$ be branches such that\/
  $\Clx=\Dlx$. Then\/ $\GMod$ satisfies\/ $\Cl$ if and only if\/ $\GMod$ satisfies\/ $\Dl$.\qed
\end{prop}

\begin{prop}[Soundness]
  Let $\Dl_1,\ldots,\Dl_n$ be the branches obtained from a branch $\Cl$ by a
  rule of\/ $\Calc$. Then $\Cl$ is
  satisfiable if and only if there is some $i\in\{1,\ldots,n\}$ such that
  $\Dl_i$ is satisfiable.\qed
\end{prop}

\begin{figure}[t]
  \begin{mathpar}
    \inferrule*[left=$\Rand$]{(s\mand t)x}{s x,~t x}
    \and
    \inferrule*[left=$\Ror$]{(s\mor t)x}{s x~~|~~t x}
    \and
    \inferrule*[left=$\Rdmnd$,
    right=\mbox{$Y\textup{ fresh}$,~$|Y|=n+1$, $\Dmnd{r}{n}t x$ \textup{not quasi-evident on} $\Cl$}]
    {\Dmnd{r}{n}t x}{\forall y\in Y\colon~r x y,~t y,~\forall z\in Y, y\ne z\colon~y\dneq z}
    \and
    \inferrule*[left=$\Rbox$,
    right=\mbox{$Y\subseteq\{y\,|\,r x y\in\Clx\}$,~$|Y|=|Y/_{\sim_\Cl}|=n+1$}]
    {\Bx{r}{n}t x}{\exists y,z\in Y,~y\ne z\colon~y\deq z~~|~~\exists y\in Y\colon~t y}
    \and
    \inferrule*[left=$\Rem$,right=\mbox{$Y\textup{ fresh}$,~$|Y|=n+1$, $\EM_n t x$ \textup{not evident on} $\Cl$}]
    {\EM_n t x}{\forall y\in Y\colon~t y,~\forall z\in Y,~y\ne z\colon~y\dneq z}
    \and
    \inferrule*[left=$\Rum$,right=$\mbox{$Y\subseteq\Nam\Cl,~|Y|=|Y/_{\sim_\Cl}|=n+1$}$]
    {\UM_n t x}{\exists y,z\in Y,~y\ne z\colon~y\deq z~~|~~\exists y\in Y\colon~t y}
    \\
    \inferrule*[left=$\RN$]{\nm{x}y}{x\deq y}
    \and
    \inferrule*[left=$\RNn$]{\mneg\nm{x}y}{x\dneq y}
    \and
    \inferrule*[left=$\Rneg$,right=\mbox{$p x\in\Clx$}]
    {\mneg p x}{\bot}
    \and
    \inferrule*[left=$\RNnI$,right=\mbox{$x\sim_\Cl y$}]{x\dneq y}{\bot}
  \end{mathpar}~\\
  {\small
  \begin{center}
    $\Cl$ is the branch to which a rule is applied.\\
    ``$Y\textup{ fresh}$'' stands for $Y\!\cap\Nam\Cl=\emptyset$.
  \end{center}
  }
  \caption{Tableau rules for $\Calc$}
  \label{BasicTableauRules}
\end{figure}

\begin{exa}
Consider the unsatisfiable formula $(\Dmnd{r}{1}p\mand\Bx{r}{1}\mneg p)x$.
Applied to the formula, our tableau rules produce three closed branches as
shown in Fig.~\ref{fig:closed-example-basic}. All the rule applications except
$\Rbox$ produce exactly one extension. The rule $\Rbox$ applies to the
formula $\Bx{r}{1}\mneg p x$ and the set $Y=\{y,z\}$ producing three
extensions. The leftmost branch is closed with $\RNnI$ applied to $y\dneq z$,
the other two branches are closed with $\Rneg$ applied to the respective two
formulas introduced by the application of $\Rbox$.
Note that without the restriction that the equational closures of alternative
extensions must be different the application of $\Rbox$ would introduce an
additional fourth extension, namely by the equation $z\deq y$.
\end{exa}

\begin{figure}
  \renewcommand{\arraystretch}{1.5}
  \centering
  \[
  \begin{array}{cr@{\quad}|@{\quad}cr@{\quad}|@{\quad}cr}
    \multicolumn{5}{c}{(\Dmnd{r}{1}p\mand\Bx{r}{1}\mneg p)x} & \\
    \multicolumn{5}{c}{\Dmnd{r}{1}p x,~\Bx{r}{1}\mneg p x} & \Rand\\
    \multicolumn{5}{c}{r x y,~p y,~r x z,~p z,~y\dneq z} & \Rdmnd\\
    \hline
    y\deq z & \Rbox & \mneg p y & \Rbox & \mneg p z & \Rbox\\
    \bot & \RNnI & \bot & \Rneg & \bot & \Rneg
  \end{array}
  \]
  \caption{Tableau derivation for $(\Dmnd{r}{1}p\mand\Bx{r}{1}\mneg p)x$}
  \label{fig:closed-example-basic}
\end{figure}

\subsection{Control} \label{sec:control}
The restrictions on the applicability of the tableau rules given by the
evidence conditions are not sufficient for termination.
Consider $\Cl_0:=\{\UM_0\Dmnd{r}{0}p x\}$. An application of
$\Rum$ to $\Cl_0$ yields $\Cl_1:=\Cl_0\cup\{\Dmnd{r}{0}p x\}$, which
can be extended by $\Rdmnd$ to $\Cl_2:=\Cl_1\cup\{r x y,\,p y\}$. Now
$\Rum$ is applicable again and yields $\Cl_3:=\Cl_2\cup\{\Dmnd{r}{0}p y\}$,
which in turn can be extended by $\Rdmnd$, and so ad infinitum.

To obtain a terminating
calculus, the rule $\Rdmnd$ needs to be restricted further.
We do so by weakening the notion of evidence for diamond formulas. The weaker
notion, called quasi-evidence, is then used in the side condition of $\Rdmnd$
in place of evidence.
As we have mentioned before, an evident branch contains a complete
description of a model of all of its formulas. A quasi-evident branch
will contain only a partial description of such a model.
In particular, quasi-evidence will not require that for every diamond
$\Dmnd{r}{n}{t}x$, we have $n+1$ outgoing edges $r x y$.
However, we require that the partial description given by a quasi-evident
branch can always be completed to a full model of the branch by adding edges.
So, in particular, every quasi-evident branch will be satisfiable.
In the above example, $\Cl_3$ will turn out to be quasi-evident and hence
terminal. And indeed, $\Cl_3$ is clearly satisfiable and can be
completed to an evident branch by adding the edge $r y y$.

While quasi-evidence was introduced in the context of pattern-based blocking,
it can also be made sense of in the context of chain-based blocking. Unlike
with pattern-based blocking, calculi using chain-based blocking usually
terminate with branches that are not quasi-evident, which is due to the presence
of ``blocked'' parts, i.e., parts of the branch that have at some point been
identified as irrelevant for the model construction and so have been excluded
from further processing.
The parts that are not blocked form a kernel from which a model can be
constructed. And in many cases, this kernel is precisely what we call a
quasi-evident branch. A concrete example relating chain-based blocking and
quasi-evidence is given in~\cite{KaminskiSmolkaJoLLI}.

Our task is now to define a notion of quasi-evidence that is weak enough
to guarantee termination of our
calculus but strong enough to preserve completeness in the presence of graded
modalities.
The notions of quasi-evidence used in previous work on pattern-based
blocking~\cite{KaminskiSmolkaM4M07,KaminskiSmolkaJoLLI} turn out to be too weak.
For instance, intuitively adapting the
notion in~\cite{KaminskiSmolkaM4M07} would give us the following candidate
definition:

A formula $\Dmnd{r}{m}s x$ is quasi-evident on $\Cl$ if there are
nominals $y,z_1,\ldots,z_{m+1}$ such
that $\{r y z_1,s z_1,\ldots,r y z_{m+1},s z_{m+1}\}\subseteq\Clx$ and
$\{\Bx{r}{n}t y\,|\,\Bx{r}{n}t x\in\Clx\}\subseteq\Clx$. (We also say:
$\Dmnd{r}{m}s x$ is quasi-evident if the corresponding \emph{pattern}
$\{\Dmnd{r}{m}s\}\cup\{\Bx{r}{n}t\,|\,\Bx{r}{n}t x\in\Clx\}$ is
\emph{expanded}).

With this definition of quasi-evidence,
no rule of our calculus would apply to the following branch:
\[\Cl:=\{r y z,~q z,~\Bx{r}{1}(p\mand\mneg p)y,~\Dmnd{r}{0}q x,~\Bx{r}{1}(p\mand\mneg p)x,~r x u,~\mneg q u\}\]
As $\Cl$ is clearly unsatisfiable, the notion of quasi-evidence needs to be
adapted.

Given a branch $\Cl$ and a role $r$,
an \emph{$r$-pattern}
is a set of expressions of the form $\mu s$, where
$\mu\in\{\Dmnd{r}{n},\Bx{r}{n}\,|\,n\in\NN\}$.
We write $\PatCl{r}{x}$ for the largest $r$-pattern $P$
such that $P\subseteq\{t\,|\,t x\in\Clx\}$. We call $\PatCl{r}{x}$ the
$r$-pattern of $x$ on $\Cl$. An $r$-pattern $P$ is \emph{expanded on $\Cl$} if
there are nominals $x,y$
such that $r x y\in\Clx$ and $P\subseteq\PatCl{r}{x}$.
In this case, we say that the nominal $x$ \emph{expands\/ $P$ on\/ $\Cl$}.

A diamond formula $\Dmnd{r}{n}s x\in\Cl$ is \emph{quasi-evident on $\Cl$} if it
is either evident on $\Cl$ or $x$ has no \emph{$r$-successor}
on $\Cl$
(i.e., there is no
$y$ such that
$r x y\in\Clx$)
and $\PatCl{r}{x}$ is expanded on $\Cl$.
The rule $\Rdmnd$ can only be applied to diamond formulas that are
not quasi-evident.

Note that whenever $\Dmnd{r}{n}s x\in\Cl$ is quasi-evident but not evident on
$\Cl$, there is a nominal $y$ that expands $\PatCl{r}{x}$ on $\Cl$.

We call a branch $\Cl$ quasi-evident if it satisfies all of the evidence
conditions but the one for diamond formulas, which we replace by:
\begin{align*}
  \Dmnd{r}{n}t x\in\Cl
  &~\Rightarrow~ \Dmnd{r}{n}t x\textup{ is quasi-evident on }\Cl
\end{align*}

\begin{exa} \label{exa:q-ev-basic}
Figure~\ref{fig:example-q-evident-basic} shows a tableau derivation
resulting in a quasi-evident branch. Let us write $\Gamma_n$ for the
branch obtained in line $n$ of the derivation. Note that
$\PatCli{\Cl_3}{r}{x}=\{\Dmnd{r}{0}p,\,\Dmnd{r}{0}q\}$ is expanded on
$\Cl_3$.
The notion of expandedness
is such that, once expanded, a pattern
remains expanded on all extensions of the branch. In particular,
if $\PatCli{\Cl_i}{r}{x}$ is expanded on $\Cl_i$, then
$\PatCli{\Cl_i}{r}{x}$ (not, however, $\PatCli{\Cl_j}{r}{x}$) will be
expanded on $\Cl_j$ for all $j\ge i$.
Note that the pattern of a nominal may change over time, i.e.,
$\PatCli{\Cl_i}{r}{x}$ and $\PatCli{\Cl_j}{r}{x}$ may be different
if $i\ne j$.
So, in the example,
$\PatCli{\Cl_1}{r}{x}=\emptyset\subsetneq\PatCli{\Cl_3}{r}{x}$.
In general, we have
$\PatCli{\Cl_i}{r}{x}\subseteq\PatCli{\Cl_j}{r}{x}$ whenever $i\le j$.
However, if $x\sim_{\Cl_i}y$ and $x$ expands $\PatCli{\Cl_i}{r}{y}$ on
$\Cl_i$, then $x$ will expand $\PatCli{\Cl_j}{r}{y}$ on $\Cl_j$ for all
$j\ge i$.

Since $\PatCli{\Cl_5}{r}{x}=\PatCli{\Cl_5}{r}{y}$, $\PatCli{\Cl_5}{r}{y}$
is expanded on $\Cl_5$, and hence both $\Dmnd{r}{0}p y$ and $\Dmnd{r}{0}q y$
are quasi-evident on $\Cl_5$. The pattern
$\PatCli{\Cl_5}{r'}{y}=\{\Dmnd{r'}{0}q\}$ is not expanded on $\Cl_5$, so
$\Rdmnd$ is applicable to $\Dmnd{r'}{0}q y$. On the branch $\Cl_6$ resulting
from this application, the pattern becomes expanded, and so does
$\PatCli{\Cl_6}{r'}{x}$. The only diamond formula that is not quasi-evident
on $\Cl_6$ is $\Dmnd{r}{0}q x$ (since it is not evident and $x$ has a successor
on $\Cl_6$). After applying $\Rdmnd$ to $\Dmnd{r}{0}q x$, $\Cl_7$ contains
only quasi-evident diamond formulas. To make the branch evident, it remains
to propagate the universal constraint
$\Dmnd{r}{0}p\mand\Dmnd{r}{0}q\mand\Dmnd{r'}{0}q$ to $z$ and $u$ (steps 8-11).
Since this introduces no new patterns (we have
$\PatCli{\Cl_{11}}{r}{z}=\PatCli{\Cl_{11}}{r}{u}=\PatCli{\Cl_{11}}{r}{y}=\PatCli{\Cl_{11}}{r}{x}$
and $\PatCli{\Cl_{11}}{r'}{z}=\PatCli{\Cl_{11}}{r'}{u}=\PatCli{\Cl_{11}}{r'}{x}=\PatCli{\Cl_{11}}{r'}{y}$),
$\Cl_{11}$ is quasi-evident.
\end{exa}

\begin{figure}
  \renewcommand{\arraystretch}{1.3}
  \centering
  \[
  \begin{array}{lc@{\qquad}l}
0. & \UM_0(\Dmnd{r}{0}p\mand\Dmnd{r}{0}q\mand\Dmnd{r'}{0}q)x &\\[.1cm]
1. & (\Dmnd{r}{0}p\mand\Dmnd{r}{0}q\mand\Dmnd{r'}{0}q)x & \Rum\\[.1cm]
2. & (\Dmnd{r}{0}p\mand\Dmnd{r}{0}q)x,~\Dmnd{r}{0}p x,~\Dmnd{r}{0}q x,
    ~\Dmnd{r'}{0}q x & 2\times\Rand\\[.1cm]
3. & r x y,~p y & \Rdmnd\\[.1cm]
4. & (\Dmnd{r}{0}p\mand\Dmnd{r}{0}q\mand\Dmnd{r'}{0}q)y & \Rum\\[.1cm]
5. & (\Dmnd{r}{0}p\mand\Dmnd{r}{0}q)y,~\Dmnd{r}{0}p y,~\Dmnd{r}{0}q y,
    ~\Dmnd{r'}{0}q y & 2\times\Rand\\[.1cm]
6. & r' y z,~q z & \Rdmnd\\[.1cm]
7. & r x u,~q u & \Rdmnd\\[.1cm]
8. & (\Dmnd{r}{0}p\mand\Dmnd{r}{0}q\mand\Dmnd{r'}{0}q)z & \Rum\\[.1cm]
9. & (\Dmnd{r}{0}p\mand\Dmnd{r}{0}q)z,~\Dmnd{r}{0}p z,~\Dmnd{r}{0}q z,
    ~\Dmnd{r'}{0}q z & 2\times\Rand\\[.1cm]
10. & (\Dmnd{r}{0}p\mand\Dmnd{r}{0}q\mand\Dmnd{r'}{0}q)u & \Rum\\[.1cm]
11. & (\Dmnd{r}{0}p\mand\Dmnd{r}{0}q)u,~\Dmnd{r}{0}p u,~\Dmnd{r}{0}q u,
    ~\Dmnd{r'}{0}q u & 2\times\Rand
  \end{array}
  \]
  \caption{Tableau derivation for $\UM_0(\Dmnd{r}{0}p\mand\Dmnd{r}{0}q\mand\Dmnd{r'}{0}q)x$}
  \label{fig:example-q-evident-basic}
\end{figure}

\begin{lem} \label{lem:q-evident-to-evident}
  Let $\Cl$ be a quasi-evident branch and let $\Dmnd{r}{n}s x\in\Cl$ be
  not evident on
  $\Cl$. Let $y$ be a nominal that expands $\PatCl{r}{x}$ on $\Cl$ and let
  $\Dl:=\Cl\cup\{r x z\,|\,r y z\in\Clx\}$.
  Then:
  \begin{enumerate}[\em(1)]
  \item
    $\forall z:~
    r x z\in\Dlx~\Longleftrightarrow~r y z\in\Clx$,
  \item $\forall m,t:~\Dmnd{r}{m}t\in\PatCl{r}{x}~\Longrightarrow~
    \Dmnd{r}{m}t x$ evident on $\Dl$,
  \item $\Dmnd{r}{n}s x$ evident on $\Dl$,
  \item $\forall r',m,t,z:~\Dmnd{r'}{m}t z$ evident on $\Cl$
    $\Longrightarrow$ $\Dmnd{r'}{m}t z$
    evident on $\Dl$,
  \item $\Dl$ quasi-evident.
  \end{enumerate}
\end{lem}

\proof 
  We begin with (1). Let $z$ be a nominal. By construction, it holds
  $r y z\in\Clx~\Rightarrow~r x z\in\Dl$. The converse implication holds by the
  fact that $\Dmnd{r}{n}s x$ is quasi-evident but not evident on $\Cl$, meaning
  that $x$ has no $r$-successor on $\Cl$.
  It remains to show:
  $r x z\in\Dl~\Leftrightarrow~r x z\in\Dlx$. The direction from left to right
  is obvious. For the other direction, assume $r x z\in\Dlx$. Then there are
  $x',z'$ such that $x'\sim_\Cl x$, $z'\sim_\Cl z$, and $r x' z'\in\Dl$. Since
  $x$ has no $r$-successor on $\Cl$,
  neither does $x'$. Hence, by the definition of $\Dl$,
  we must have $x'=x$, and so $r x z'\in\Dl$. But then $r y z'\in\Clx$, and
  consequently $r y z\in\Clx$. The claim follows by
  the definition of $\Dl$.

  Now to (2). Let $\Dmnd{r}{m}t\in\PatCl{r}{x}$. Since
  $\PatCl{r}{y}\supseteq\PatCl{r}{x}$, in
  particular it holds $\Dmnd{r}{m}t y\in\Clx$, i.e., there is some
  $y'\sim_\Cl y$ such that $\Dmnd{r}{m}t y'\in\Cl$. By (1), it suffices
  to show that $\Dmnd{r}{m}t y$ is evident on $\Cl$. This is the case
  since $\Dmnd{r}{m}t y'$ is quasi-evident on $\Cl$ (as $\Cl$ is
  quasi-evident) and $y'$ has an $r$-successor on $\Cl$ (as $y$ has one on
  $\Cl$).

  Claim (3) immediately follows from (2).

  Claim (4) is obvious as the evidence of diamonds on a branch cannot be
  destroyed by adding edges.

  Now to (5). The only conditions that might in principle be violated on $\Dl$
  are the quasi-evidence condition for diamonds of the form
  $\Dmnd{r}{m}t z\in\Dl$ where $z\sim_\Dl x$,
  and the evidence condition for boxes
  $\Bx{r}{m}t z\in\Dl$ where $z\sim_\Dl x$.

  For diamonds of the above form,
  the quasi-evidence condition holds by (2).

  If $\Bx{r}{m}t z\in\Dl$ and $z\sim_\Dl x$,
  it holds $\Bx{r}{m}t y\in\Clx$ since
  $\PatCl{r}{y}\supseteq\PatCl{r}{x}=\PatDl{r}{x}$. Hence by (1) it suffices to
  show that $\Bx{r}{m}t y$ is evident on $\Cl$, which is the case since $\Cl$ is
  quasi-evident.
\qed

\begin{thm}[Evidence Completion] \label{thm:evidence-completion}
  For every quasi-evident branch $\Cl$ there is an evident branch $\Dl$ such
  that $\Cl\subseteq\Dl$.
\end{thm}

\proof 
  For every branch $\Cl$ we define:
  \[\varphi\Cl:=|\{\Dmnd{r}{n}s x\,|\,\Dmnd{r}{n}s x\in\Cl\land
  \Dmnd{t}{n}s x\textup{ not evident on }\Cl\}|\]
  Let $\Cl$ be quasi-evident. We proceed by induction on
  $\varphi\Cl$.
  If $\varphi\Cl=0$, then $\Cl$ is evident and we are done. Otherwise, there is
  a diamond $\Dmnd{r}{n}s x\in\Cl$ that is not evident on $\Cl$.
  Let $y$ be a nominal that expands $\PatCl{r}{x}$ on $\Cl$, and
  let $\Cl':=\Cl\cup\{r x z\,|\,r y z\in\Clx\}$. By
  Lemma~\ref{lem:q-evident-to-evident}(3-5), $\Cl'$ is quasi-evident and
  $\varphi\Cl'<\varphi\Cl$. So, by the inductive hypothesis, there is some
  evident branch $\Dl$ such that
  $\Cl\subseteq\Cl'\subseteq\Dl$.
\qed

A branch is called \emph{maximal} if it cannot be extended by
any tableau rule.

\begin{thm}[Quasi-evidence] \label{thm:max-to-q-evident}
  Every open and maximal branch in\/ $\Calc$ is quasi-ev\-i\-dent.
\end{thm}

\proof 
  Let $\Cl$ be an open and maximal branch. Note that we have no evidence
  or quasi-evidence conditions for formulas of the form $p x$, $r x y$ or
  $x\deq y$.
  We show that every $s\in\Cl$ that is not of the form $p x$, $r x y$ or
  $x\deq y$
  is (quasi-)evident on $\Cl$ by case analysis on the shape of $s$.
  \begin{enumerate}[$\bullet$]
  \item $s=\mneg p x$. The claim, $p x\notin\Clx$, follows by $\Rneg$
    (and the assumption that $\Cl$ is open and maximal).
  \item $s=x\dneq y$. The claim, $x\not\sim_\Cl y$, follows by $\RNnI$
    (and the assumption that $\Cl$ is open and maximal).
  \item $s=\nm x y$. 
    By $\RN$, $x\deq y\in\Clx$ and hence $x\sim_\Cl y$.
  \item $s=\mneg\nm x y$.
    By $\RNn$, $x\dneq y\in\Clx$.
    Then there are some $x'$ and $y'$ such that $x'\sim_\Cl x$, $y'\sim_\Cl y$,
    and $x'\dneq y'\in\Cl$. By $\RNnI$, we have $x'\not\sim_\Cl y'$
    (cf. $s=x\dneq y$). The claim follows by the transitivity of $\sim_\Cl$.
  \item $s=\Bx{r}{n}t x$. To show:
    $|\{y\,|\,r x y\in\Clx,~t y\notin\Clx\}/_{\sim_\Cl}|\le n$. This is clearly
    the case if $|\{y\,|\,r x y\in\Clx\}|\le n$. Otherwise, it suffices to show
    that for every $Y\subseteq\{y\,|\,r x y\in\Clx\}$ such that $|Y|=n+1$,
    it either holds $|Y/_{\sim_\Cl}|<|Y|$ or $t y\in\Clx$ for some $y\in Y$.
    This follows by $\Rbox$
    since $y\deq z\in\Cl$ implies $y\sim_\Cl z$ for all $y,z\in Y$.
  \end{enumerate}
  The cases $s=(t_1\mor t_2)x$, $s=(t_1\mand t_2)x$, and $s=\Dmnd{r}{n}t x$
  are immediate by, respectively, $\Ror$, $\Rand$, and $\Rdmnd$.
  The
  cases $s=\EM_n t x$ and $s=\UM_n t x$ are proved analogously to
  $s=\Dmnd{r}{n}t x$ and, respectively, $s=\Bx{r}{n}t x$.
\qed

\subsection{Termination}
We will now show that every tableau derivation is finite. As usual, the main
difficulty is bounding the number of applications of generative rules, in
particular of $\Rdmnd$. The present proof is notably more complex than the
proofs
in~\cite{KaminskiSmolkaM4M07,KaminskiSmolkaJoLLI} since now, an application
of $\Rdmnd$ does not necessarily expand a new pattern. Hence, we need to
combine the pattern-counting argument
from~\cite{KaminskiSmolkaM4M07,KaminskiSmolkaJoLLI} with a bound on the number
of non-expanding applications of $\Rdmnd$.

Since the rules
$\Ror$, $\Rbox$, and $\Rum$ are all finitely branching, by
König's lemma it suffices to show that the construction of every individual
branch terminates. Since tableau rule application always produces
proper extensions of branches, it then suffices to show that the size (i.e.,
cardinality) of an individual branch is bounded.

First, we show that the size of a branch $\Cl$ is bounded by a function in the
number of nominals on $\Cl$. Then, we show that this number itself is bounded
from above, completing the termination proof.

We write $\Cl\ext{\mathcal{R}}\Dl$ to denote that the branch $\Dl$ is
obtained from $\Cl$ by the rule $\mathcal{R}$. We write $\Cl\to\Dl$ if $\Dl$ is
obtained from $\Cl$ by a single rule application.
We write $\STer\Cl$ for the set of all modal expressions occurring on $\Cl$,
possibly as subterms of other expressions, and $\Rel\Cl$ for the set of all
roles that occur on $\Cl$.

Crucial for the termination argument is the fact the tableau rules cannot
introduce any modal expressions that do not already occur
on the initial branch.

\begin{prop} \label{prop:subterm-prop}
  If $\Cl,\Dl$ are branches such that $\Dl$ is obtained from $\Cl$ by any rule
  of $\Calc$,
  then $\STer\Dl=\STer\Cl$.\qed
\end{prop}

For every pair of nominals $x,y$ a branch $\Cl$ may contain an equation
$x\deq y$ or a disequation $x\dneq y$. For every pair $x,y$ and every role $r$,
$\Cl$ may contain an edge $r x y$. Moreover, for every expression
$s\in\STer\Cl$, $\Cl$ may contain a
formula $s x$. Hence, the size of $\Cl$ is bounded by
$(2+|\Rel\Cl|)\cdot|\Nam\Cl|^2+|\STer\Cl|\cdot|\Nam\Cl|$.
By Proposition~\ref{prop:subterm-prop}, we know that $|\STer\Cl|$ and
$|\Rel\Cl|$ depend only on the initial branch.
Clearly, $|\STer\Cl|$ and $|\Rel\Cl|$ are bounded from above by the size
of the input, i.e., the sum of the sizes of the initial formulas.

By the above,
it suffices to show that $|\Nam\Cl|$ is exponentially bounded in the size of
the input. We do so by
giving a bound on the number of applications of $\Rdmnd$ and $\Rem$ that can
occur in the derivation of a branch, which suffices since $\Rdmnd$ and $\Rem$
are the only two rules that can introduce new nominals.

We begin by showing that $\Rem$ can be applied at most as many times as there
are distinct modal
expressions of the form $\EM_n s$ on the initial branch.
For this purpose, we define a function $\fE$ such that
$\fE\Cl:=\{\EM_n s\in\STer\Cl\,|\,\exists x\in\Nam\Cl:~\EM_n s x$ not evident
on $\Cl\}$.
Since $|\fE\Cl|$ is bounded from below by $0$, it suffices to show that the
number decreases with every application of $\Rem$ (and is
non-increasing otherwise, which is obvious).

\begin{lem} \label{lem:evidence-preservation}
  Let $s$ be of the form $\Dmnd{r}{n}t x$ or $\EM_n t x$. If $s$ is evident on\/
  $\Cl$ and\/ $\Cl\subseteq\Dl$, then $s$ is evident on $\Dl$.\qed
\end{lem}

\begin{prop} \label{prop:em-exp-terminates}
  $\Cl\ext{\Rem}\Dl~\Longrightarrow~|\fE\Cl|>|\fE\Dl|$
\end{prop}

\proof 
  Let $\Cl\ext{\Rem}\Dl$. By Lemma~\ref{lem:evidence-preservation},
  $\fE\Cl\supseteq\fE\Dl$. Hence it suffices to show that $\fE\Cl-\fE\Dl$ is
  non-empty. Let $\Dl$ be obtained from $\Cl$ by applying $\Rem$ to
  $s=\EM_n t x$. Then, by $\Rem$, $\EM_n t\in\fE\Cl$. On the other hand, $s$
  is evident on $\Dl$, and it is easy to see that the evidence of $s$ implies
  the evidence of $\EM_n t y$ for every $y\in\Nam\Dl$. Hence
  $\EM_n t\notin\fE\Dl$.
\qed

Now we show that $\Rdmnd$ can be applied at most finitely often in a derivation.
Since there are only finitely many roles, it suffices to show
that $\Rdmnd$ can be applied at most finitely often for each role.
Observe that since $\Rdmnd$ is only applicable to
diamond formulas that are not quasi-evident,
it holds:

\begin{prop} \label{prop:dmnd-applicability-cases}
  If\/ $\Rdmnd$ is applicable to a formula $\Dmnd{r}{n}s x\in\Cl$, then either
  \begin{enumerate}[\em(1)]
  \item $x$ has an $r$-successor on\/ $\Cl$, or
  \item $\PatCl{r}{x}$ is not expanded on\/ $\Cl$.\qed
  \end{enumerate}
\end{prop}
Let $\Cl$ and $\Dl$ be branches such that $\Dl$ is obtained from $\Cl$ by
applying $\Rdmnd$ to a formula $\Dmnd{r}{n}s x\in\Cl$ such that
$\PatCl{r}{x}$ is not expanded on $\Cl$. It is easy to see that $\PatDl{r}{x}$
must be expanded on $\Dl$. Let us call such an application of $\Rdmnd$
\emph{pattern-expanding}.

Let
$\Pat^r\Cl:=\mathcal{P}(\{\Dmnd{r}{n}s\in\STer\Cl\}\cup\{\Bx{r}{n}s\in\STer\Cl\})$.
In other words, $\Pat^r\Cl$ contains all the possible sets of $r$-diamonds and
$r$-boxes from $\STer\Cl$.
Since $\Cl\to\Dl$ implies $\Clx\subseteq\Dlx$, it holds:

\begin{lem} \label{lem:pat-preservation}
Let $\Cl\to\Dl$ and $P\in\Pat^r\Cl$. If $P$ is expanded on $\Cl$, then $P$ is
expanded on~$\Dl$.\qed
\end{lem}
So, for each role $r$
the derivation
of a branch has at most $|\Pat^r\Cl_0|$ pattern-expanding applications of
$\Rdmnd$, where $\Cl_0$ is the initial branch. Clearly, $|\Pat^r\Cl_0|$ is
exponentially bounded in the size of the input.

Hence, it remains to show that a derivation can contain only finitely many
applications of $\Rdmnd$ assuming that none of the applications is
pattern-expanding.
We say a nominal $x$ has a \emph{successor} on $\Cl$ if $x$ has an $r$-successor
on $\Cl$ for any role $r$.
A set of nominals $X$ has
a successor on $\Cl$ if there is some $x\in X$ that has a successor on $\Cl$.
We define \[\fD^{X}\Cl:=|\{\Dmnd{r}{n}s\in\STer\Cl\,|\,\exists x\in X\colon~
\Dmnd{r}{n}s x\textup{ not evident on }\Cl\}|\] and
\[\fD\Cl:=\sum_{\begin{subarray}{c}X\in\Nam\Cl/_{\sim_\Cl}\\
X\textup{ has a successor on }\Cl\end{subarray}}\fD^{X}\Cl\enspace.\]

\begin{lem} \label{lem:succ-preservation}
  Let $X,Y\in\Nam\Cl/_{\sim_\Cl}$, $x\in X$, $y\in Y$, and let
  $\Dl:=\Cl\cup\{x\deq y\}$. Then
  $\fD^X\Cl\ge\fD^{X\cup Y}\Dl$.\qed
\end{lem}

\begin{prop} \label{prop:dmnd-exp-terminates}
  Let $\Cl\to\Dl$
  such that $\Dl$ is obtained from $\Cl$ by some rule application other than
  a pattern-expanding application of $\Rdmnd$.
  \begin{enumerate}[\em(1)]
  \item If $\Dl$ is obtained from $\Cl$ by $\Rdmnd$, then
    $\fD\Cl>\fD\Dl$.
  \item Otherwise, $\fD\Cl\ge\fD\Dl$.
  \end{enumerate}
\end{prop}

\proof\hfill
  \begin{enumerate}[(1)]
  \item Clearly, nominals introduced by $\Rdmnd$ are fresh and hence cannot
    have any successors on $\Dl$. Hence $\fD\Cl\ge\fD\Dl$. Therefore, it
    suffices to find a set $X\in\Nam\Cl/_{\sim_\Cl}$
    that has a successor
    on $\Cl$, a nominal $x\in X$ and a
    formula $\Dmnd{r}{n}s x\in\Cl$ that is not evident on $\Cl$ but is evident
    on $\Dl$.

    Assume $\Dl$ is obtained from $\Cl$ by $\Rdmnd$ applied to a formula
    $\Dmnd{r}{n}s x\in\Cl$. Clearly, $\Dmnd{r}{n}s x$ is not evident on $\Cl$
    but is evident on $\Dl$. Since the rule application is not
    pattern-expanding, $x$ has an $r$-successor on $\Cl$. Hence there is some
    $X\in\Nam\Cl/_{\sim_\Cl}$ such that $x\in X$ and $X$ has a successor on
    $\Cl$. The claim follows.
  \item Since cumulativity of tableau construction preserves the evidence of
    diamond formulas (Lemma~\ref{lem:evidence-preservation}), the only interesting rules are those modifying
    $\Nam\Cl/_{\sim_\Cl}$. Nominals introduced by $\Rem$ are fresh and hence do
    not have any successors on $\Dl$.
    Therefore, the only remaining cases are $\RN$,
    $\Rbox$ and $\Rum$. Clearly,
    none of the three rules can increase the cardinality of
    $\{X\in\Nam\Cl/_{\sim_\Cl}\,|\,X\textup{ has a successor on }\Cl\}$.
    The claim follows by Lemma~\ref{lem:succ-preservation}.\qed
  \end{enumerate}

This completes the termination proof. Since the cardinalities of the sets
$\Pat^{r}\Cl$ are exponentially bounded in the size $n_0$ of the input,
$|\fE\Cl|$ is polynomial in $n_0$,
and $\fD\Cl$ polynomial
in $|\Cl|$ and $n_0$, $|\Nam\Cl|$ is exponentially bounded in $n_0$.
Since $|\Cl|$ is
polynomial in $|\Nam\Cl|$, we conclude that $|\Cl|$ is at most exponential in
$n_0$.
By cumulativity, the construction of $\Cl$ terminates in at most exponentially
many steps in $n_0$.
This
suffices to give us
a \textsc{NExpTime} complexity bound for
the decision procedure based on the calculus.

\section{Adding Reflexivity, Transitivity and Role Inclusion}
We now extend $\Calc$ to deal with reflexivity, transitivity and inclusion
assertions. As in related work on description
logic~\cite{HorrocksPHD,HorrocksSattlerTobies00,HorrocksSattler01,HorrocksSattler07,HorrocksEtAl06}, we
restrict our modal expressions to contain no graded boxes for roles that have
transitive subroles.

We define $\subseteq^*_\Cl$ as the smallest reflexive and transitive
relation such that \mbox{$r\subseteq^*_\Cl r'$} whenever $r\subsetax r'\in\Cl$.
A role $r$ is called \emph{simple} on a branch $\Cl$ (or just simple if $\Cl$
is clear from the context) if there is no $r'$ such
that $r'\subseteq^*_\Cl r$ and $\Trans r'\in\Cl$.
Observe that all subroles of a simple role are in turn simple.
Also, since our tableau rules will not introduce new inclusion assertions,
a role $r$ will be simple on a given branch $\Cl$ if and only if $r$ is simple
on the initial branch from which $\Cl$ is obtained.

Our branches may now contain inclusion, reflexivity
and transitivity assertions:
\begin{align*}
  s &~::=~ t x\;|\;r x y\;|\;x\deq y\;|\;x\dneq y\;|\;\bot\;|\;r\subsetax r'
  \;|\;\Ref r\;|\;\Trans r
\end{align*}
The modal expressions $t$ in formulas of the form $t x$ are restricted to
contain no boxes $\Bx{r}{n}s$ with $n>0$ unless $r$ is simple.

Following the ideas
in~\cite{HorrocksPHD,HorrocksSattler01,HorrocksSattler07,HorrocksEtAl06}, we
introduce
the \emph{induced transition relation}\/~$\AccRef{r}{}{}$ to reason about
accessibility in the presence of inclusion axioms. Intuitively,
$\AccRef{r}{x}{y}$ means that in every model of $\Cl$, $y$ is accessible
from $x$ via~$r$.

\subsection{Extending Evidence}
To account for the new types of formulas, we extend the evidence conditions
as follows:
\begin{align*}
  r\subsetax r'\in\Cl &~\Rightarrow~
  \forall x,y\in\Nam\Cl:~r x y\in\Clx\,\Rightarrow\, r' x y\in\Clx\\
  \Ref r\in\Cl &~\Rightarrow~
  \forall x\in\Nam\Cl:~r x x\in\Clx\\
  \Trans r\in\Cl &~\Rightarrow~
  \forall x,y,z\in\Nam\Cl:~r x y\in\Clx\,\land\, r y z\in\Clx\,\Rightarrow\,
  r x z\in\Clx
\end{align*}
It is easy to see that if $\Cl$ satisfies the extended evidence conditions,
the interpretation $\ModCl$ constructed in the proof of
Theorem~\ref{thm:evident-branches-sat} will satisfy the new formulas. Hence,
Theorem~\ref{thm:evident-branches-sat} adapts to the extended
system.

\begin{thm}[Model Existence] \label{thm:evident-branches-sat-full}
  Every evident branch has a finite model.\qed
\end{thm}

\subsection{Pre-evidence} \label{sec:pre-evidence-x}
To account for the new evidence conditions, one could imagine the following
rules.
\begin{mathpar}
  \inferrule*{r\subsetax r',~r x y}{r' x y}
  \and
  \inferrule*[right=$x\in\Nam\Cl$]{\Ref r}{r x x}
  \and
  \inferrule*[]{\Trans r,~r x y,~r y z}{r x z}
\end{mathpar}
In the presence of blocking,
however, the rules are problematic. In particular, the
rule for reflexivity renders the notion of quasi-evidence that we use for
$\Calc$
ineffective to ensure termination. Once we add a reflexive edge $r x x$ to a
branch
$\Cl$, $x$ will have an $r$-successor on~$\Cl$, meaning quasi-evidence will
coincide with evidence for all $r$-diamonds on $x$. Similarly, the rule for
transitivity is
known to be incomplete in the presence of blocking~\cite{KaminskiSmolkaJoLLI}.

We solve the problem by defining a weaker notion of evidence,
called \emph{pre-evidence}. To satisfy the pre-evidence conditions, we do not
have to explicitly add reflexive or transitive edges during tableau
construction.
We will extend our tableau rules and the notion of quasi-evidence such that
every open and maximal branch in the extended calculus can
be completed to a pre-evident branch,
which in turn can be made evident by adding the implicit edges.

We define the relation $\Acc{r}{}{}$ as the least relation such that:
\begin{eqnarray*}
r x y\in\Clx &~\Rightarrow~& \Acc{r}{x}{y}\\
r'\subsetax r\in\Cl,~\Acc{r'}{x}{y} &~\Rightarrow~& \Acc{r}{x}{y}
\end{eqnarray*}
The relation $\Acc{r}{}{}$ does not account for reflexivity. To do so,
we extend it as follows:
\begin{align*}
\AccRef{r}{}{} &~:=~\left\{
  \begin{array}{l@{\quad}l}
    \Acc{r}{}{}\cup\,\{(x,y)\,|\,x,y\in\Nam\Cl\land x\sim_\Cl y\} & \textup{if }\exists r':~r'\subseteq^*_\Cl r\land\Ref r'\in\Cl\\
    \Acc{r}{}{} & \textup{otherwise}
  \end{array}\right.
\end{align*}

The \emph{pre-evidence conditions} are obtained from the evidence conditions by
omitting the conditions for inclusion and reflexivity assertions and
replacing the conditions for diamonds, boxes and
transitivity assertions as follows:
\begin{align*}
  \Dmnd{r}{n}t x\in\Cl
  &~\Rightarrow~
  \exists^{n+1} Y\!:~D_\Cl Y~\land~\forall
  y\in Y:~
  \AccRef{r}{x}{y}~\land~ t y\in\Clx\\
  \Bx{r}{n}t x\in\Cl
  &~\Rightarrow~ |\{y\,|\,\AccRef{r}{x}{y},~t y\notin\Clx\}/_{\sim_\Cl}|\le n\\
  \Trans r\in\Cl &~\Rightarrow~ \forall r',t,x,y:~
  \Bx{r'}{0}t x\in\Clx~\land~r\subseteq^*_\Cl r'~\land~\Acc{r}{x}{y}~\Rightarrow~\Bx{r}{0}t y\in\Clx
\end{align*}
Note that we do not need pre-evidence conditions for inclusion or
reflexivity assertions as their semantics is taken care of by the way we define
the relation $\AccRef{r}{x}{y}$.
Pre-evidence of individual formulas is defined analogously to the corresponding
notion of evidence.

We now show that every pre-evident branch can be extended to an evident branch.
Let the \emph{evidence closure} $\evClo{\Cl}$ of a branch $\Cl$ be defined as
the least superset of $\Cl$ such that:
  \begin{align*}
    \AccRef{r}{x}{y} &~\Rightarrow~ r x y\in\evClo{\Cl}\\
    \Trans r\in\Cl~\land~ r x y\in\evClo{\Cl}~\land~ r y z\in\evClo{\Cl}
    &~\Rightarrow~ r x z\in\evClo{\Cl}\\
    r\subsetax r'\in\Cl~\land~ r x y\in\evClo{\Cl}
    &~\Rightarrow~ r' x y\in\evClo{\Cl}
  \end{align*}
Note that, by construction, we have
$r x y\in\evClo{{\Clx}}\iff r x y\in\evClo{\Cl}$.

\begin{lem} \label{lem:ev-closure-simple}
  Let $\Cl$ be a branch and\/ $r$ be simple on $\Cl$. Then
  $\AccRef{r}{x}{y}\iff r x y\in\evClo{\Cl}$
\end{lem}

\proof 
  Let $r$ be simple on $\Cl$.
  The direction from left to right is immediate.
  The other direction
  can be shown by induction on the construction of $\evClo{\Cl}$ from $\Cl$.
\qed

\begin{lem} \label{lem:trans-chains}
  Let $\Cl$ be a branch and let $r x y\in\evClo{\Cl}$. Then either
  $\AccRef{r}{x}{y}$, or
  there is some $r'$ such that $\{r'\subsetax r,\,\Trans r'\}\subseteq\Cl$ and
  \[\exists\,n{\ge}2\,\exists x_1,\ldots,x_n\!:~x_1=x\,\land\,x_n=y\,\land
  \,\forall 1{\le}i{<}n:~\Acc{r'}{x_i}{x_{i+1}}\enspace.\]
\end{lem}

\proof 
  By induction on the construction of $\evClo{\Cl}$.
\qed

\begin{thm}[Evidence Completion]
  $\Cl$ pre-evident $\Longrightarrow$ $\evClo{\Cl}$ evident
\end{thm}

\proof 
  It is easy to see that $\evClo{\Cl}$ satisfies the evidence conditions for
  inclusion, reflexivity and transitivity assertions.
  The only remaining evidence conditions that may be affected by adding edges to
  $\Cl$ are the ones for diamonds and boxes. The rest of the evidence conditions
  is already satisfied by $\Cl$ and hence also holds on~$\evClo{\Cl}$.

  The evidence condition for diamonds holds on $\evClo{\Cl}$ since the
  corresponding pre-evidence condition holds on $\Cl$ and
  $\AccRef{r}{x}{y}$ implies $r x y\in\evClo{\Cl}$ for all nominals $x,y$
  and roles $r$.

  It remains to show the evidence condition for boxes. Let $\Bx{r}{n}s x\in\Cl$
  and $|\{y\,|\,{\AccRef{r}{x}{y}},$ $s y\notin\Clx\}/_{\sim_\Cl}|\le n$. It
  suffices to show:
  $|\{y\,|\,r x y\in\evClo{\Cl},~s y\notin\Clx\}/_{\sim_\Cl}|\le n$.
  We distinguish two cases. If $r$ is simple, the claim follows by
  Lemma~\ref{lem:ev-closure-simple}. Otherwise, we must have $n=0$. Hence, it
  suffices to show that we have $s y\in\Clx$ for every edge
  $r x y\in\evClo{\Cl}$. Let $r x y\in\evClo{\Cl}$. Then, by
  Lemma~\ref{lem:trans-chains}, two cases are possible. Either
  $\AccRef{r}{x}{y}$, in which case the
  claim follows by the pre-evidence condition for boxes, or there is a
  transitive subrole $r'$ of $r$ such that there are nominals
  $x_1,\ldots,x_m$ ($m\ge 2$) such that $x_1=x$, $x_m=y$ and
  $\Acc{r'}{x_i}{x_{i+1}}$ for all $1\le i<n$.
  In this case, by induction on $m$ one can show that the pre-evidence
  condition for transitivity assertions applied to $r'$ and $\Bx{r}{n}s x\in\Cl$
  implies either $\Bx{r}{n}s x_{m-1}\in\Cl$ (true by assumption for $m=2$) or
  $\Bx{r'}{n}s x_{m-1}\in\Clx$ (if $m>2$).
  Either way, the claim follows by the pre-evidence condition for boxes.
\qed

\subsection{Tableau Rules}
The tableau rules for the extended calculus $\CalcX$ in Fig.~\ref{TableauRules}
replace
the original rule
$\Rbox$ from Fig.~\ref{BasicTableauRules} and
add a new rule $\Rt$, which is necessary to achieve the pre-evidence condition
for transitivity assertions.
While the formulation of $\Rdmnd$ remains unchanged, the rule will now have to
use
an adapted notion of quasi-evidence, which
will be introduced in Sect.~\ref{sec:control-x}. For now,
we assume $\Rdmnd$ is formulated with
the restriction ``$\Dmnd{r}{n}t x$ not pre-evident on $\Cl$'' instead.
Again, it is not hard to verify that the extended rules are sound.

\begin{figure}[t]
  \begin{mathpar}
    \inferrule*[left=$\Rbox$,
    right=\mbox{$Y\subseteq\{y\,|\,\AccRef{r}{x}{y}\}$,~$|Y|=|Y/_{\sim_\Cl}|=n+1$}]
    {\Bx{r}{n}t x}{\exists y,z\in Y,~y\ne z\colon~y\deq z~~|~~\exists y\in Y\colon~t y}
    \and
    \inferrule*[left=$\Rt$,right=\mbox{$r\subseteq^*_\Cl r',~\Acc{r}{x}{y}$}]
    {\Trans r,~\Bx{r'}{0}t x}{\Bx{r}{0}t y}
  \end{mathpar}
  \caption{New rules for $\CalcX$}
  \label{TableauRules}
\end{figure}

\subsection{Control} \label{sec:control-x}
As it turns out, in the presence of role inclusion we have to modify the
definition of patterns. It no longer suffices to consider patterns
separately for each role.
This is due to the fact that now, different roles may be constrained
by inclusion assertions.
Consider, for instance, the unsatisfiable branch
\[\Cl:=\{r\subsetax r',~\Dmnd{r}{0} p x,~\Dmnd{r'}{0}\mneg p x,
~\Bx{r'}{1}(p\mand\mneg p)x,~r' x y,~\mneg p y,~\Dmnd{r}{0}p z,~r z u,~p u\}\]
According to our previous notion of quasi-evidence, $\Dmnd{r}{0}p x$ is
quasi-evident on $\Cl$ as $x$ has no $r$-successor (even if we extend the set
of successors to $\{y\,|\,\Acc{r}{x}{y}\}$) and
$\PatCl{r}{x}$ is expanded. Since the other two diamonds on $\Cl$ are evident,
$\Cl$ is quasi-evident, witnessing the incompleteness of our previous
definition of patterns.

Hence, we redefine the notion of a pattern as follows.
Given a branch $\Cl$, a \emph{pattern}
is a set of terms of the form $\mu s$, where
$\mu\in\{\Dmnd{r}{n},\Bx{r}{n}\,|\,r\in\Rel\Cl,~n\in\NN\}$.
We write $\UPatCl{x}$ for the largest pattern $P$
such that $P\subseteq\{t\,|\,t x\in\Clx\}$. We call $\UPatCl{x}$ the
pattern of $x$ on $\Cl$. A pattern $P$ is \emph{expanded on $\Cl$} if
there are nominals $x,y$ and a role $r$ such that
$\Acc{r}{x}{y}$
and $P\subseteq\UPatCl{x}$.
In this case, we say that
$x$ \emph{expands\/ $P$ on\/ $\Cl$}.
Note that here we use the relation $\Acc{r}{}{}$ rather than~$\AccRef{r}{}{}$.
Otherwise, we would get the same problems with termination as outlined in
Sect.~\ref{sec:pre-evidence-x}.

A diamond formula $\Dmnd{r}{n}s x$ is \emph{quasi-evident on $\Cl$} if it is
either pre-evident on $\Cl$ or $x$ has no
\emph{successor} on $\Cl$ (i.e., there is no $y$ and $r$ such that
$\Acc{r}{x}{y}$)
and $\UPatCl{x}$ is expanded on $\Cl$.
As before, we restrict the rule $\Rdmnd$ such that it can only be applied to
diamond formulas that are not quasi-evident, and
call a branch $\Cl$ quasi-evident if it satisfies all of the pre-evidence
conditions but the one for diamond formulas, which we again replace by
\begin{align*}
  \Dmnd{r}{n}t x\in\Cl
  &~\Rightarrow~ \Dmnd{r}{n}t x\textup{ is quasi-evident on }\Cl
\end{align*}
but now with the adapted notion of quasi-evidence.

\begin{exa}
  Figure~\ref{fig:ev-tableau-x} shows a tableau derivation in
  $\CalcX$ resulting in a quasi-evident branch.
  As in Example~\ref{exa:q-ev-basic}, we write $\Cl_n$ for the branch
  up to line $n$. We observe:
  \begin{enumerate}[$\bullet$]
  \item Since $r$ is reflexive and $r\subsetax r'\in\Cl_0$, $r'$ is also
    reflexive. Consequently, we have $\AccRefCli{\Cl_0}{r'}{x}{x}$, which
    explains why $\Rbox$ applies to $\Bx{r'}{0}\Dmnd{r}{0}p x\in\Cl_0$.
  \item The rule $\Rt$ propagates $\Bx{r'}{0}\Dmnd{r}{0}p$ to $z$ but not
    to $y$ since $r$ is not (necessarily) transitive.
  \item In $\Calc$, $\Dmnd{r}{0}p z\in\Cl_5$ would be
    quasi-evident since
    $\PatCli{\Cl_5}{r}{x}=\PatCli{\Cl_5}{r}{z}$. In
    $\CalcX$, however, $\Rdmnd$ applies to $\Dmnd{r}{0}p z$ since
    ${\PatCli{\Cl_5}{}{x}=\{\Dmnd{r}{0}p,\,\Dmnd{r'}{0}q,\,\Bx{r'}{0}\Dmnd{r}{0}p\}\ne\{\Dmnd{r}{0}p,\,\Bx{r'}{0}\Dmnd{r}{0}p\}=\PatCli{\Cl_5}{}{z}}$.
  \end{enumerate}
\end{exa}

\begin{figure}
  \renewcommand{\arraystretch}{1.3}
  \centering
  \[
  \begin{array}{lc@{\qquad}l}
    0.&r\subsetax r',~\Ref r,~\Trans r',
    ~\Bx{r'}{0}\Dmnd{r}{0}p x,~\Dmnd{r'}{0}q x
    &\\
    1.&\Dmnd{r}{0}p x & \Rbox\\
    2.&r x y,~p y & \Rdmnd\\
    3.&r' x z,~q z & \Rdmnd\\
    4.&\Bx{r'}{0}\Dmnd{r}{0}p z & \Rt\\
    5.&\Dmnd{r}{0}p z & \Rbox\\
    6.&r z u,~p u & \Rdmnd
  \end{array}
  \]
  \caption{Tableau derivation for $\{r\subsetax r',~\Ref r,~\Trans r',~\Bx{r'}{0}\Dmnd{r}{0}p x,~\Dmnd{r'}{0}q x\}$}
  \label{fig:ev-tableau-x}
\end{figure}

\begin{lem} \label{lem:acc-ind-prop}
  Let $\Cl,\Dl$ be branches such that
  $\{r\sqsubseteq r'\,|\,r\sqsubseteq r'\in\Cl\}
  =\{r\sqsubseteq r'\,|\,r\sqsubseteq {r'\in\Dl}\}$.
  Let $x$, $y$, $u$, $v$ be nominals such that
  $\{r\,|\,r x y\in\Clx\}=\{r\,|\,r u v\in\Dlx\}$. Then, for all
  $r$, ${\Acc{r}{x}{y}}~\Leftrightarrow~\AccDl{r}{u}{v}$.
\end{lem}

\proof
  Let $\Cl$, $\Dl$, $x$, $y$, $u$ and $v$ be as required. Let $r$ be a role.
  We show
  ${\Acc{r}{x}{y}}~\Rightarrow~\AccDl{r}{u}{v}$ by induction on the
  derivation of ${\Acc{r}{x}{y}}$. The other direction follows analogously
  by induction on the derivation of ${\AccDl{r}{u}{v}}$.
  Assume ${\Acc{r}{x}{y}}$. We distinguish two cases:
  \begin{enumerate}[$\bullet$]
  \item $r x y\in\Clx$. Then, by assumption, $r u v\in\Dlx$, and so
    ${\AccDl{r}{u}{v}}$.
  \item There is some $r'$ such that $r'\sqsubseteq r\in\Cl$ and
    ${\Acc{r'}{x}{y}}$. By the inductive hypothesis, we have
    ${\AccDl{r'}{u}{v}}$. Moreover, by assumption, $r'\sqsubseteq r\in\Dl$.
    Hence, ${\AccDl{r}{u}{v}}$. \qed
  \end{enumerate}

\begin{lem} \label{lem:q-evident-to-pre-evident}
  Let $\Cl$ be a quasi-evident branch and let
  $\Dmnd{r}{n}s x$ be not
  pre-evident on $\Cl$. Let $y$
  expand $\UPatCl{x}$ on $\Cl$
  and let
  $\Dl:=\Cl\cup\{r' x z\,|\,r' y z\in\Clx\}$.
  Then:
  \begin{enumerate}[\em(1)]
  \item
    $\forall r',z:~
    \AccDl{r'}{x}{z}~\Longleftrightarrow~\Acc{r'}{y}{z}\;$
    and
    $~\AccRefDl{r'}{x}{z}~\Longleftrightarrow~\AccRef{r'}{y}{z}$,
  \item $\forall r',m,t:~
    \Dmnd{r'}{m}t\in\UPatCl{x}~\Longrightarrow~\Dmnd{r'}{m}t x$ pre-evident
    on $\Dl$,
  \item $\Dmnd{r}{n}s x$ pre-evident on $\Dl$,
  \item $\forall r',m,t,z:~\Dmnd{r'}{m}t z$ pre-evident on
    $\Cl~\Longrightarrow~\Dmnd{r'}{m}t z$ pre-evident on $\Dl$,
  \item $\Dl$ quasi-evident.
  \end{enumerate}
\end{lem}

\proof 
  We begin with (1).
  Let $r'$ be a role and $z$ a nominal. We will only show
  the first equivalence
  since the other claim easily
  follows.
  Since $\Dmnd{r}{n}s x$ is quasi-evident but not evident on $\Cl$, $x$ has no
  successor on $\Cl$. Hence, by construction,
  $\{r'\,|\,r' x z\in\Dlx\}=\{r'\,|\,r' y z\in\Clx\}$. The claim follows by
  Lemma~\ref{lem:acc-ind-prop}.

  Claims (2--4) are shown analogously to the corresponding claims of
  Lemma~\ref{lem:q-evident-to-evident}.

  Now to (5). The only conditions that might in principle be violated in $\Dl$
  are the quasi-evidence condition for diamonds of the form
  $\Dmnd{r'}{m}t z\in\Dl$ where $z\sim_\Dl x$,
  the evidence condition for boxes $\Bx{r'}{m}t z\in\Dl$ where $z\sim_\Dl x$,
  and the evidence condition for transitivity
  assertions $\Trans r'\in\Dl$.

  For diamonds of the above form, the quasi-evidence condition holds
  by (2).

  For transitivity assertions, it suffices to show that for every
  $r_1,r_2$ such that
  $\Trans r_1\in\Cl$, $r_1\subseteq^*_\Cl r_2$, and $\Bx{r_2}{0}t x\in\Clx$, and
  for all $z$ such that $\AccDl{r_1}{x}{z}$, it holds $\Bx{r_1}{0}t z\in\Clx$.
  Since $\UPatCl{y}\supseteq\UPatCl{x}$, we have $\Bx{r_2}{0}t y\in\Clx$.
  The claim now follows by (1) and the quasi-evidence condition for
  $\Trans r_1\in\Cl$.

  The claim for boxes follows analogously
  (we exploit $\UPatCl{y}\supseteq\UPatCl{x}$ and (1)).
\qed

\begin{thm}[Pre-evidence Completion]
  For every quasi-evident branch $\Cl$ there is a pre-evident branch $\Dl$ such
  that $\Cl\subseteq\Dl$.
\end{thm}

\proof 
  Proceeds analogously to the proof of Theorem~\ref{thm:evidence-completion}
  with Lemma~\ref{lem:q-evident-to-pre-evident} in place of
  Lemma~\ref{lem:q-evident-to-evident}.
\qed

\begin{thm}[Quasi-evidence] \label{thm:max-to-q-evident-x}
  Every open and maximal branch in\/ $\CalcX$ is qua\-si-evident.
\end{thm}

\proof 
  Proceeds analogously to the proof of Theorem~\ref{thm:max-to-q-evident}.
  The additional case for transitivity assertions is straightforward.
\qed

\subsection{Termination}

The termination proof for $\CalcX$ proceeds analogously to the proof for
$\Calc$. Let us sketch what needs to be adapted.
Because of the rule $\Rt$, the set $\STer\Cl$ of modal expressions occurring on
$\Cl$ needs to be extended as follows: $\STerNew\Cl:=\STer\Cl\cup
\{\Bx{r}{0}s\,|\,r\subseteq^*_\Cl r'\land\Bx{r'}{0}s\in\STer\Cl\}$.
With the extended definition of $\STer$,
Proposition~\ref{prop:subterm-prop}
holds for $\CalcX$.
Lemma~\ref{lem:evidence-preservation} is modified as follows:
\begin{lem}
  Let $s$ be of the form $\Dmnd{r}{n}t x$ or $\EM_n t x$. If $s$ is
  (pre-)evident on $\Cl$ and $\Cl\subseteq\Dl$, then $s$ is (pre-)evident on
  $\Dl$.\qed
\end{lem}
Proposition~\ref{prop:em-exp-terminates} is unaffected by the extensions to the
calculus. Proposition~\ref{prop:dmnd-applicability-cases} is adapted as follows:
\begin{prop}
  If\/ $\Rdmnd$ is applicable to a formula\/ $\Dmnd{r}{n}s x\in\Cl$, then either
  \begin{enumerate}[\em(1)]
  \item $x$ has a successor on $\Cl$, or
  \item $\UPatCl{x}$ is not expanded on $\Cl$.\qed
  \end{enumerate}
\end{prop}
Also, analogously to Lemma~\ref{lem:pat-preservation}, the expandedness of our
extended patterns is preserved by tableau rule application.
Lemma~\ref{lem:succ-preservation} and Proposition~\ref{prop:dmnd-exp-terminates}
remain valid if we
redefine
\[\fD^{X}\Cl:=|\{\Dmnd{r}{n}s\in\STerNew\Cl\,|\,\exists x\in X\colon~
\Dmnd{r}{n}s x\textup{ not pre-evident on }\Cl\}|\] and
$\fD\Cl$ accordingly, with the modified definition of a successor.

\section{Conclusion}
We have presented a terminating tableau calculus for graded hybrid logic with
global modalities and role hierarchies.
Following~\cite{BolanderBrauner06,BolanderBlackburn07,KaminskiSmolkaJoLLI}, our
calculus is cumulative, representing state equality abstractly via an
equivalence relation (declarative approach). The existing calculi for equivalent
and
stronger logics~\cite{HorrocksSattler01,HorrocksSattler07,HorrocksEtAl06}
work on possibly cyclic graph structures and treat equality by destructive graph
transformation during tableau construction (procedural approach). The procedural
approach encompasses algorithmic decisions that are not present in the more
abstract declarative approach. From a declarative calculus we can always
obtain a procedural system by refinement.

Exploiting an extended pattern-based blocking technique and the cumulativity of
our calculus, we have proved a $\textsc{NExpTime}$ complexity bound for the
associated decision procedure.
To ensure termination of pattern-based blocking in the presence of reflexivity,
we differentiated between the induced transition relation
$\AccRef{r}{}{}$ and
its non-reflexive counterpart $\Acc{r}{}{}$.
The implementation of pattern-based blocking for
a hybrid language with global modalities~\cite{GoetzmannEtAl:2009:Spartacus}
reveals its considerable practical potential. We consider it a promising
project to implement the extended version of pattern-based blocking presented
in this paper and compare its performance to that of established blocking
techniques.

Following related
work~\cite{HorrocksPHD,HorrocksSattlerTobies00,HorrocksSattler01,HorrocksSattler07,HorrocksEtAl06},
we restrict the language decided by our calculus to
contain no graded boxes on complex roles. As shown by Horrocks, Sattler and
Tobies~\cite{HorrocksSattlerTobies00}, this restriction is essential for
decidability of logics extending $\mathcal{SHIN}$. In the absence of inverse
roles ($\mathcal{I}$), however, the restriction of graded boxes to
simple roles can be significantly relaxed~\cite{KazakovEtAl07}.
In~\cite{KaminskiSmolkaTCS10}, we give a terminating tableau calculus for
$\mathcal{SOQ}$ extended by graded boxes on transitive roles.
The logic extends the decidable fragment of~\cite{KazakovEtAl07} by nominals
but lacks inclusion assertions that are allowed (with some restrictions)
in~\cite{KazakovEtAl07}.
It remains an open problem to design an efficient tableau calculus for the full
decidable fragment of~\cite{KazakovEtAl07}.
Also, it is still open if the fragment of~\cite{KazakovEtAl07} remains
decidable when extended by nominals.

\subsection*{Acknowledgement}
We would like to thank our referees for their
valuable comments that helped to improve the paper.

\bibliography{paper_pfe}
\bibliographystyle{plain}

\end{document}